# Time irreversibility in the quantum systems with infinite number of particles

Yuping Huo*

*School of Physics, Zhengzhou University, Zhengzhou, 450001 Henan, China*

Abstract

The time irreversible evolution of quantum systems with infinite number of particles (QSINP) was studied within a newly constructed algebraic framework. The QSINP could be described by the quantum infinite lattice field. The *-Algebra R is the set of all dynamic variables of the QSINP, in which a special addition and multiplication operations were defined. The full pure state vector (FPSV) ρ, the pure state vector $ρ_f$ and the equivalent relations within them are also well defined. The set of all pure state vectors, which are equivalent to the FPSV ρ, is a Hilbert Space $H_ρ$, if the addition, the inner product and the norm were defined in it. The set of all linear transformation $N_ρ$ on $H_ρ$ is isomorphic to R, and $\{N_ρ, H_ρ\}$ is the representation of R, associated with ρ. The *-Algebra R has infinitely many non-equivalent irreducible GNS constructions (representations), associated with different nonequivalent FPSVs respectively.

It is proved that，the dynamical motions of the QSINP is totally within the GNS construction associated with the initial pure state vector; However, the time reversal transformation makes the initial FPSV and its corresponding dynamical evolution into another non-equivalent GNS construction. Therefore, within the GNS construction associated with the FPSV, the dynamics of the QSINP is time irreversible.

Finally, due to the Liouville operator has real continuous spectrum within the GNS construction, the longtime asymptotic solution of Liouville equation in the GNS construction could be treated by the formal scattering theory. The Master equation with dissipative term was obtained, which is formally irrelevant of the initial state and the corresponding GNS construction. This master equation could be regarded as the evolution equation of the QSINP on R.

\* Email: huoyp@zzu.edu.cn    (Y. P. Huo)

**I. Introduction**

The dynamical basis of the time irreversibility of many-particle systems, which marks the microscopic origin of the second law of thermodynamics, has always been a long-standing unresolved problem. For more than hundred years many scientists had studied this problem from different theoretical methods, such as, the dynamic system theory (including the Ergodic theory, mixing system theory, non-integrable system), the chaotic orbits of non-linear oscillators, the chaotic behavior of non-linear systems and so no [1-4]. But none of them can elucidate why almost all macroscopic systems are dissipative (time irreversible).

This is why some people could not believe that, the time reversibility of dynamics of finitely many particles could be changed to time irreversible only by the thermodynamic limit processes. To meet this, some specific "statistical assumptions" was introduced into many-bodies theory. The typical examples are the "coarse grain" processes and the Bogolubov theory with its assumption [4], etc. In these cases, the results of *approaching equilibrium* could really be reached. But the physical basis, or the meaning of such kind of assumptions, is still unclear as well as the macroscopic time irreversibility itself. Therefore, up to now no one has claimed that, the final solution of the microscopic origin of the second law of thermodynamics does to be obtained. On the other hand, many scientists think that it is best to discuss this unsolved problem by directly analyzing the evolution of the quantum system with infinite number of particles (QSINP). Such suggestion also has been supported by the Haag theorem in quantum field theory [5-7].

By using quantum field theory, an almost completely framework has been developed for the quantum statistical mechanics, which including: the Fock space of *quantum system of many particles*: $H_f = \oplus_n H_1^n$ ; second quantization; the set of linear operators on $H_F$; the Liouville equation: $dA/dt = -iLA$ , or the time evolution operator $\text{Exp}(-iLt)$; the Green's function Hierarchy and different cut-off procedures; partial summation of infinite perturbation series in the interaction representation; etc., where $H_1$ is the Hilbert space of single particle, $H_1^n$ is the Hilbert space of *n* particles, L is the Liouville operator, and A is a dynamical variable associated with finitely many particles. However, the Fock space contains infinite products of $H_1$, it is non-Hilbert space. How to analyze the structure of such non-Hilbert space and treat the operators on it, especially the Liouville operator, are still the unresolved problems. Even if the Fock space $H_F$ was discussed under the framework of thermodynamic limit, that is to say, the Fock space $H_F$ was regarded as a Hilbert space, the above theory cannot solve the problems associated with spontaneously breaking of the time reversal invariance [8].

Ever since the thirties of last century, much effort are made to describe the systems with infinitely many degrees of freedom by C*-algebra theory [9-10]. However, the $C^*$-algebra was defined as *-Banach algebra at the very beginning. Such theory is only suitable to apply to some special problems, for instance, the non-interacting-particle systems or the KMS state in equilibrium statistical mechanics [9-10]. This maybe the reason why the $C^*$-algebra theory has not been widely used by physicists nowadays.

In this paper, the spontaneously breaking of the *time reversal invariance* in the non-relativistic quantum system with infinite number of particles (QSINP) is investigated. The paper contains three parts. (i) The algebraic framework for the dynamics of the SQINP is well developed. The quantum lattice field on infinite space has been used to describe the QSINP, its dynamical variables are generated by the grid point operators (creation and destruction operators at each grid point) at finitely many grid points. The *-Algebra R is the set of all dynamic variables of the QSINP, in which special addition and multiplication operations were defined. The full pure state vector (FPSV) ρ, the pure state vector $ρ_f$ and the equivalent relations within them are also defined. The set of all pure state vectors, which are equivalent to the FPSV ρ, is a Hilbert Space $H_ρ$, if the addition, the inner product and the norm were defined in it. Even though the norm or metric cannot be defined either on R, it can be proved that, from a fully pure state vector ρ, an irreducible representation of R, $\{N_ρ, H_ρ\}$ (GNS Construction) can be constructed, where $H_ρ$ is the Hilbert space composed of a set of mutually equivalent pure state vectors, $H_\rho$ is the set of all linear operators on $H_\rho$ (isomorphic to R). The *-Algebra R has infinitely many non-equivalent irreducible GNS constructions associated with different non-equivalent set of fully pure state vectors respectively. In contrast, according to the Stone-von Neumann theorem, the C*-Algebra, which is composed of the dynamical variables of the systems with finite number of particles, has only one unique non-equivalent irreducible representation. (ii) Since $N_\rho$ is isomorphic to R, the Liouville operator L defined on R is also a linear Hermitian operator on $N_\rho$. The Liouville equation only describe the motion of the elements of $N_\rho$ in $N_\rho$, so any dynamical process of the QSINP is totally inside the GNS Construction $\{N_\rho, H_\rho\}$ associated with its initial pure state vector ρ. But, the time reversal state $\rho_T$ of ρ, is the element of another Hilbert space $H_{\rho T}$, the GNS constructions $\{N_{\rho T}, H_{\rho T}\}$ is non-equivalent to $\{N_\rho, H_\rho\}$. Hence the dynamics of the QSINP is time irreversible. (iii) Since the Liouville operator on $N_\rho$ has real continuous spectrum, the long-time asymptotic solution of the Liouville equation can be derived by the formal scattering theory, which is an operator of simi-group with a dissipative generator. The corresponding master equation does not formally depend on ρ or $\{N_\rho, H_\rho\}$, so it could be thought as the dissipative evolution equation on R, and can be treated approximately by the traditional methods in quantum field theory or quantum many-body theory.

**II  The *-algebra of dynamical variables and its local representations**

Since the density of the QSINP must be finite everywhere and the infinitely many particles have to be distributed in infinite space accordingly. This is to say, there is no boundary of the system in any finite spatial region. In principle, the QSINP can be described by the complex field function $\psi(x)$ which is smoothly distributed over the whole infinite space. It belongs to the space $C_0$, instead of the Hilbert space. As we known, there is no effective mathematical tool to deal with such $C_0$-kind of quantum field up to now. Due to the Heisenberg uncertainty principle (each particle should not be considered as an infinitesimal point in the space), it is natural to describe the QSINP by the infinite lattice field $\psi(I)$, which is similar to the case of modern quantum field theory. Here the grid point is marked by a set of three integers $I=\{i_1, i_2, i_3\}$. All the grid points compose the infinite series of grid points N. The grid spacing $\Delta x$ should be small enough. The second quantization of boson lattice field is,

$$\left[\psi'(I'), \psi(I)\right] = \Delta x^3 \delta_{I',I} \tag{1}$$

$$\left[\psi(I'), \psi(I)\right] = 0. \tag{2}$$

The particle number operator on grid point I is,

$$N(I) = \psi^*(I)\psi(I). \tag{3}$$

The fermionic field satisfies the anti-commutation relation.

Because the lattice field operators at different grid points are commutable, each point is an independent degree of freedom, and each degree of freedom has a couple of conjugate operators $\{\psi^*(I), \psi(I)\}$. It is also possible to rearrange the set of three integers $\{i_1, i_2, i_3\}$ as one positive integer series $\{k\}$, k=0, 1, 2,…, from which the QSINP can be treated as a quantum system with countable and denumerable infinite degrees of freedom and the algebraic theory can be formulated regularly.

It is worth noting that, describing the QSINP by the lattice field in infinite space is necessary, or at least a much advantageous choice, because of the following reasons: (i) The number of particles in any finite spatial region must be finite. (ii) Any singlet particle should occupy a finite size space from the point of view of quantum field theory. Traditionally, in quantum field theory, the regulation procedure should be applied to eliminate the ultra-violet divergence. (iii) The dynamics of the whole system should be translation invariance so long as the displacement is an integer number of lattice spacing.

First of all, we need to develop the algebraic framework for the dynamics of the SQINP.

For each singlet grid point "I", all the dynamical variables, generated with the grid point operators $\{\psi^*(I), \psi(I)\}$, compose the C*-Algebra $R(I)$. According to the Stone-von Neumann theorem, $R(I)$ only has one unique non-equivalent irreducible representation $\{N(I), H(I)\}$, here $H(I)$ is the Hilbert space. The C*-Algebra $N(I)$ is the set of all linear operators on $H(I)$, and it is isomorphic to $R(I)$. In $H(I)$, there is a complete orthogonal basis: $\{\varphi(I, n)\}_n$, which satisfies,

$$\psi(I)\varphi(I, 0)=0,$$

$$\psi^*(I)\psi(I)\varphi(I, n)=n\varphi(I, n),$$

$$(\varphi(I, n), \varphi(I, n))=1, n=0,1,2,\ldots \quad (4)$$

Any one element in $H(I)$ is called the grid state vector $\varphi(I)$, which can be expanded by $\{\varphi(I, n)\}_n$,

$$\varphi(I) = \sum_{n=1} C_n \varphi(I, n) \quad (5)$$

According to the C*-Algebra theory, the state $\xi(I)$ on the grid point I is defined as the positive linear functional on $R(I)$, denoted as: $\{(A(I)), \xi(I) \mid \text{for all } A(I) \in R(I)\}$, which satisfies,

$$\left(A^*(I)A(I), \xi(I)\right) \geq 0. \quad (6)$$

The state space $P_I$ composed of all states $\xi(I)$ on $R(I)$, is a convex set, in which the extreme state $\xi^P(I)$ is determined by the normalized grid point state vector $\varphi(I)$ in $H(I)$,

$$\left(A(I), \xi^P(I)\right) = \left(\varphi(I)^*, A(I)\varphi(I)\right), \quad A(I) \in N(I) \text{ and } \varphi(I) \in H(I). \quad (7)$$

For any two grid point state vectors $\varphi_1(I)$ and $\varphi_2(I)$, there exists an operator $A_{12}(I)$ in $R(I)$ (or $N(I)$) to satisfy $\varphi_1 = A_{12}(I)\varphi_2(I)$. At each grid point, the grid point unit operator $E(I)$ has also been defined: $E(I)E(I)=E(I)$; $E(I)A(I)=A(I)$.

For the QSINP, only the dynamic variables of the sub-systems with finitely many degrees of freedom are finite and have real physical meaning. The set R of all dynamic variables of the QSINP should be composed of the dynamic variables of all the possible sub-systems with different finitely many degrees of freedom. In other words, the dynamic variable of the QSINP has to be defined as the algebraic combination of the grid operators $\{\psi^*(I), \psi(I)\}_I$ at finitely many grid points [9-10].

Since any two grid point operators, $\psi(I)$ (or $\psi^*(I)$) and $\psi(I')$ (or $\psi^*(I')$), defined at different grid points are commutable, any dynamic variable of the QSINP $A_s$ could be written as an infinite series of grid point dynamic variables,

$$A_s = \{\{A(I)\}_{IS}, \{E(I)\}_{JS\infty}\}_N \tag{8}$$

Where $E(I)$ is the grid point unit operator at "I", $I_s$ is a finite series of grid points. $I_{s\infty}$ is the complementary series of $I_s$ in N, $A(I)$ is a grid point dynamic variable at I. The Hermitian counterpart of the element $A_s \in R$ could be defined as,

$$A_s^* = \{\{A(I)^*\}_{IS}, \{E(I)\}_{JS\infty}\}_N.$$

Because the different elements of R may have non-$E(I)$ grid point operators at different grid point series, it is necessary to redefine the algebraic operations in R. There are two different elements of R: $A_s$, $A_t \varepsilon R$, the $I_s$, $I_t$ are their non-$E(I)$ finite series of grid point separately. Suppose $I_{st}$ is the union of the series $I_s$ and $I_t$: $I_{st} = I_s \cup I_t$, $J_{st\infty}$ is the infinite complementary series of $I_{st}$ in N, the grid point operators of $A_s$ (or $A_t$) at $I_{st}$ are the same as originals: $A_s(I)$ or $E(I)$ ($A_t(I)$ or $E(I)$) respectively. The grid point operators of $A_s$, $A_t$ at $J_{st\infty}$ are all $E(I)$. Then, the addition operation is defined as,

$$A_s + A_t = 2\{\{(A_s(I) + A_t(I)/2)\}_{Ist}, \{E(I)\}_{Jst\infty}\} \tag{9}$$

and the multiplication is defined as,

$$A_s \bullet A_t = \{\{A_s(I) \bullet A_t(I)\}_{Ist}, \{E(I)\}_{Jst\infty}\}. \tag{10}$$

The above definitions of addition and multiplication satisfy the requirements of commutative and associative rules. Therefore, R is the dynamic variable *-Algebra of the QSINP. Because the QSINP has infinitely many degrees of freedom, it is impossible to define the global norm on R, R does not be C*-Algebra [11-12]. R satisfies the grid translation invariance.

With the help of "grid point state vector" at each grid point in N, it is possible to define the state vector of the QSINP. The "Fully Pure State Vector" of the QSINP (denoted by "FPSV") is an infinite series of normalized grid point state vectors at the infinite series of grid points N: $\{\varphi_n(I_n)\}_N$. Two FPSV $\rho$ and $\rho_1$ are equivalent, if $\rho_1$ has different normalized grid point state vectors with $\rho$ only at finite series of grid points. All the FPSVs, which equivalent to $\rho$ are also equivalent with each other, and compose the equivalent set of FPSV $\rho$: $D_\rho$. The Pure State Vector $\rho_f$ is an infinite series of grid point state vector (they might to be non-normalized) at the infinite series of grid points N. The $\rho_f$ is equivalent to the FPSV $\rho$, if it has only finitely many different grid point state vectors with $\rho$. All the Pure State Vectors of the QSINP, which are equivalent to the FPSV $\rho$, compose the set $H_\rho$. In other words, $\rho_f$, an element of $H_\rho$, is an infinite series of grid point state vectors at N, and has the same normalized grid point state vectors $\{\varphi_n(I)\}$ with those of $\rho$ at an infinite sub-series of grid points $J_{f\infty}$. The finite grid point series $I_f$ is the complementary series of $J_{f\infty}$ in N, at which the components of $\rho_f$ could be different with the corresponding normalized components of $\rho$. We have,

$$\rho = \{\varphi_n(I)\}_N, \qquad \rho_f = \{\{\varphi_f(I)\}_{IF}, \{\varphi_n(I)\}_{Jf\infty}\}_N \qquad (11)$$

The $H_\rho$ contains $D_\rho$. Inside $H_\rho$ it is possible to define the linear operations. Assume that, $\rho_{f1}$ and $\rho_{f2}$ are two elements of $H_\rho$, they have different components with those of $\rho$ at finite series of grid point $I_{f1}$ and $I_{f2}$ respectively, and $I_{f12}$ is the union of $I_{f1}$ with $I_{f2}$,

$$\rho_{f1} = \{\{\varphi_{f1}(I)\}_{If12}, \{\varphi_n(I)\}_{Jf12\infty}\}_N,$$

$$\rho_{f2} = \{\{f_{f2}(I)\}_{If12}, \{f_n(I)\}_{Jf12\infty}\}_N \qquad (12)$$

where $\varphi_{f1}(I)$ and $\varphi_{f2}(I)$ are the components of $\rho_{f1}$ and $\rho_{f2}$ at grid point series $I_{f12}$ respectively, $J_{f12\infty}$ is the complementary infinite series of $I_{f12}$ in N. If c and d are complex figures, $c\rho_{f1}$ is also the element of $H_\rho$, the addition operation is defined as,

$$c\rho_{f1}+d\rho_{f1}=(c+d)\{\{c\phi_{f1}(I)+d\phi_{f2}(I)/(c+d)\}_{If12},\{\phi_n(I)\}_{Jf12\infty}\}_N \quad (13)$$

So $H_\rho$ is a linear space. Using the definition of the inner-product in the grid point Hilbert Space $H(I)$, the inner-product of $\rho_{f1}$ with $\rho_{f2}$ in $H_\rho$ could be defined,

$$(\rho_{f2},\rho_{f1})=\left(\prod_{If12}(\phi_{f2}(I),\phi_{f1}(I))\right)\bullet\left(\prod_{Jf12\infty}((\phi_n(I),\phi_n(I)))\right)=\left(\prod_{If12}(\phi_{f2}(I),\phi_{f1}(I))\right) \quad (14)$$

where $(\phi_n(I),\phi_n(I))=1$. Since $I_{f12}$ is a finite series of grid point, the $(\rho_{f2},\rho_{f1})$ is a finite value.

From equation (14), the norm of the element of $H_\rho$ could be also defined,

$$|\rho_{f1}|^2=(\rho_{f1},\rho_{f1})=\left(\prod_{If1}(\phi_{1f1}(I),\phi_{f1}(I))\right) \quad (15)$$

The $H_\rho$ with inner-product (14) and norm (15) is a Hilbert Space. According to the theory of C*-Algebra, the set $N_\rho$ of all linear transformations on $H_\rho$ is a C*-Algebra.

Now the relation between the dynamic variable *-Algebra R of the QSINP and the Hilbert Space $H_\rho$ has to be established. The element of R,

$$A_s=\{\{A_s(I)\}_{IS},\{E(I)\}_{JS\infty}\}_N \quad (15')$$

has non-$E(I)$ grid point dynamic variables only at finite series of grid point $I_S$. The pure state vector $\rho_{f1}\in H_\rho$ is $\rho_f=\{\{f_f(I)\}_{If},\{f_n(I)\}_{Jf\infty}\}_N$.

Assuming the $I_{Sf}$ is the union of $I_S$ with $I_f$, $J_{Sf\infty}$ is the complimentary series of $I_{Sf}$ in N, the $A_{s\rho f}$ could be defined as,

$$A_{s\rho f}=\{\{A_s(I)f_{f1}(I)\}_{Isf},\{f_n(I)\}_{Jsf\infty}\}_N \quad (16)$$

The $A_S$ operating on $\rho_f$, only finitely many components of $\rho_f$ at grid point series $I_{Sf}$ have been changed. Thus $A_S\rho_f$ is still an element of $H_\rho$. The element of R could be thought as an linear transformation on $H_\rho$. Considering the Stone-Van Neumann theorem, the $N_\rho$ is isomorphism of R. The $\{N_\rho,H_\rho\}$ is a representation of R. Though the dynamic variable

*-Algebra R is non-measurable globally, due to its special structure, it is possible to construct a irreducible representation $\{N_\rho, H_\rho\}$ associated with a FPSV $\rho$. The $\{N_\rho, H_\rho\}$ is called the GNS Construction associated with the FPSV $\rho$.

According to the definition of FPSV, two non-equivalent FPSV's, $\rho_1$ and $\rho_2$, have different normalized grid point state vectors, $\varphi_1(I)$ and $\varphi_2(I)$, at an infinite series of grid points T. The inner-product of $\rho_1$ and $\rho_1$ is,

$$(\rho_2, \rho_1) = \prod\nolimits^{T}(\varphi_2(I), \varphi_1(I))_P \cdot \left(\prod\right)^{PT}(\phi\ \varphi( \quad (17)$$

where PT is the complimentary series of T in N, at which $\rho_1$ and $\rho_2$ have the same components. Because at any grid point in T, the two grid point state vectors: $\varphi_1(I)$ and $\varphi_2(I)$ are selected independently and $(\varphi_2(I), \varphi_1(I)) < 1$, we have, $\left(\prod\nolimits^{T} I(\varphi_2(I), \varphi_1(I))\right) = 0$, and thus,

$$(\rho_2, \rho_1) = 0 \quad (18)$$

In evidence, if $\rho_{f1}$, $\rho_{f2}$ are any elements of Hilbert spaces $H_{\rho 1}$ and $H_{\rho 2}$ separately, we also have $(\rho_{f2}, \rho_{f1}) = 0$.

The conclusion is that: given any two Hilbert spaces $H_{\rho 1}$ and $H_{\rho 2}$, associated with non-equivalent FPSV's $\rho_2$ and $\rho_1$ respectively, are mutually orthogonal globally.

The pure state vector space $\rho_f$ of the QSINP is composed of infinitely many mutually orthogonal Hilbert spaces $\{H_\rho\}$. Based on each Hilbert space $H_\rho$, which is an equivalent set of pure state vectors, which are equivalent to FPSD $\rho$, the corresponding irreducible GNS construction $\{N_\rho, H_\rho\}$ of the *-Algebra R has been constructed. The *-Algebra R has infinitely many non-equivalent irreducible GNS constructions. For comparison, unlike the *-Algebra of QSINP, the C*-Algebra of the quantum system with finite number of particles only has unique non-equivalent irreducible representation, which is the traditional framework of the quantum mechanics.

According to the general theory of algebra, the *state* $\xi$ of the QSINP should be the positive linear functional on R with some specific conditions,

$$\{(A, \xi) | (A^*A, \xi) \geq 0, A \in R\} \tag{19}$$

The set of all above defined states is complete with respect to linear combination; therefore it is the state space P of the system. Referring to the traditional quantum mechanics, from the pure state vector $\rho_f$ defined above, the corresponding pure state $\xi_f$ on R is,

$$\{(A, \xi_f) = (\rho_f, A_{\rho f}), A \in R\} \tag{20}$$

owning to $\rho_f$ is an element of the Hilbert space $H_\rho$, all the values of $(A, \xi_f)$ are finite. The pure state $\xi_f$ is a positive linear functional on R. The expectation value of a dynamic variable *a* is defined as,

$$\bar{a} = (a, \xi_f)/(\rho_f, \rho_f). \tag{21}$$

Generally, the pure state $\xi_f$ could not be expressed as the linear combination of other pure states. The *mixed state* $\xi_{mix}$ on R is defined as the positive linear combination of the pure states,

$$\xi_{mix} = \sum_\alpha f_\alpha \xi_{f\alpha}, \quad f_\alpha > 0 \tag{22}$$

$$\{(A, \xi_{mix}) | A \in R\} = \{\sum_\alpha f_\alpha (A, \xi_{f\alpha}), A \in R\} \tag{23}$$

here $\{\alpha\}$ are the labels of the pure states. Obviously, the mixed states are also positive linear functional on R. All the pure states and the mixed states compose the state space P of the QSINP. P is a convex space and all the pure states are the extreme states.

It is well known that, P is the dual space of R. Anyone element of R is generated by the grid point operators at finite grid points, but any one state $\xi$ of P involves grid states $\{\varphi(I)\}_N$ at all the infinite grid points. Such asymmetry between P and R is the main reason why the representations of R and their structures are completely different with the representations of quantum system with finite number of particles. The SQINP has infinite degrees of freedom, it is impossible to define the "norm" globally in the *-algebra R or in the state space P generally. Due to the special structure of R and the fact that the space of pure state vector $P_f$ contains infinitely many mutually orthogonal Hilbert spaces $\{H_\rho\}_\rho$, the C*-Algebras of linear operators $N_\rho$ on each of $\{H_\rho\}_\rho$ are all isomorphic to R separately. Therefore, *-Algebra R of the QSINP has infinitely many non-equivalent irreducible representations (GNS constructions $\{\{N_\rho, H_\rho\}\}_\rho$) associated with different non-equivalent FPSV's. It makes possible to develop the quantitative description of the dynamical motion of the QSINP in each GNS Construction $\{N_\rho, H_\rho\}$ associated with the corresponding initial FPSV ρ.

It has been noted that, any mixed state could be expressed uniquely by the linear sum of a set of pure states: $\xi_{mix} = \sum_\alpha f_\alpha \xi_\alpha$ with $f_\alpha > 0$, and the pure state $\xi_\alpha$ corresponds to a pure state vector $\rho_\alpha$ in the Hilbert space $H_{\rho\alpha}$ of the GNS construction $\{N_{\rho\alpha}, H_{\rho\alpha}\}$. Then we can calculate the value of $(\rho_\alpha, A\rho_\alpha)$ in $\{N_{\rho\alpha}, H_{\rho\alpha}\}$, and have,

$$(A, \xi_{mix}) = \sum_\alpha f_\alpha (\rho_\alpha, A\rho_\alpha) \tag{24}$$

We are more interesting with the semi-group $\{g_t\}$ defined on R. In order to calculate the expression of

$$(g_t A, \xi_{mix}) = \sum_\alpha f_\alpha (g_t A, \xi_\alpha) = \sum_\alpha f_\alpha (\rho_\alpha, g_t A \rho_\alpha) \tag{25}$$

we have firstly to get the corresponding GNS constructions $\{N_{\rho\alpha}, H_{\rho\alpha}\}$ associated with the pure state vectors $\rho_\alpha$, and then, make the subsequent calculations $(g_t A, \xi_\alpha)$ inside such GNS constructions. Since the operator calculations of $g_t A$ are formally independent on the concrete structure of the GNS constructions $\{N_{\rho\alpha}, H_{\rho\alpha}\}$, the final results are the mean value of $g_t A$ in mixed state $\xi_{mix}$ directly, as the usual cases in quantum statistical mechanics.

**III. Dynamics of the QSINP and Time Reversal Transformation**

Here, the dynamics of the QSINP with two-body short-range interactions will be discussed to illustrate the further potential application of the *-Algebra and its representations. For the lattice filed of the QSINP, the momentum density operator at grid point $(I=\{i1, i2, i3\})$ is,

$$P_1(I)=(\psi^*(I)(\psi(i_1+1, i_2, i_3)-\psi(i_1-1, i_2, i_3))-$$

$$(\psi^*(i_1+1, i_2, i_3)-\psi^*(i_1-1, i_2, i_3))\psi(I)))/4\Delta x \tag{26}$$

The density operator of kinetic energy is,

$$H_0(I)=\sum_{j=1}^{3}(\psi^*(I)(\psi(i_j+2)+\psi(i_j-2)-2\psi(I))+((\psi^*(i_j+2)+\psi^*(i_j-2)-2\psi^*(I))\psi(I))/8m(\Delta x)^2 \tag{27}$$

The Hamiltonian of the QSINP could be formally expressed as,

$$H=H_0+H'=\sum_I H_0(I)\Delta x^3 + \sum_I \sum_{|J-I|\leq S} V_{I,J}\Delta x^3 \tag{28}$$

where $V_{I,J}$ is the energy density operator of interaction between two particles at grid points I and J within the range of S. The summation $\sum_I$ is taken for all (infinite) grid points. The expression Hamiltonian H is the sum of the divergent series involving grid point operators at infinite grid points. Since the Hamiltonian H is not an element of R, the Schrödinger picture can't be adopted directly. But, it is still possible to formally define the Liouville operator on R,

$$LA=[i H, A]=i HA - AiH \tag{29}$$

Because any one element A in R only involves grid point operators at finite grid points, the commutator $[H, A]$ also involves only finitely many grid point operators. So the Liouville operator L is a linear operator on R even though the Hamiltonian H is a divergent series. It is easy to be proved that, L is a Hermitian operator on R and the Liouville equation is

$$dA(t)/dt = -iLA(t) \tag{30}$$

Equation (30) is the dynamic equation of the QSINP on R, which corresponding to the Heisenberg picture. The formal solution of equation (30) is

$$A(t) = T(t)A \tag{31}$$

And,

$$T(t)=\sum_{K=0}^{\infty}((-iLt)^k/k!)=\mathrm{Exp}(-iLt) \qquad (32)$$

Here, the T(t) seems to be the one parameter Lie-group on R, but in fact it is only the summation of a divergent series. In order to solve the Liouville equation quantitatively, it is necessary to know the initial value of A(t=0), i.e. to know the initial pure state vector $\rho$ of the QSINP. Obviously, for a dynamic process, $\rho$ should be a FPSV with normalized grid point state vectors at all the infinite grid points.

Since $N_\rho$ is isomorphism to R, the Liouville equation and its formal solution T(t) on R could be directly transferred onto the C*-Algebra $N_\rho$, and we will also denote the elements of $N_\rho$ as A's in R. The dynamic motion of the QSINP, described by the Liouville equation (30), is totally within the GNS construction $\{N_\rho, H_\rho\}$ associated initial FPSV $\rho$. On the other side, any element A of $N_\rho$ could also corresponds to an element $\rho_A = A_\rho$ in $H_\rho$. From the theory of GNS Construction, it is possible to presume that the elements of $N_\rho$ do not change with time and equation (30) describes the motion of the pure state vectors $A(t)\rho$ in $H_\rho$. This is similar to the Schrodinger picture in quantum mechanics. The dynamical motion of the QSINP could be also thought as being constrained totally inside the subset $H_\rho$ associated with the initial FPSV $\rho$. This conclusion is crucial for understanding the relations between the macroscopic description and the microscopic motion.

The time reversal transformation T can be introduced in R: $A \in R$, $TA=A^* \in R$. We have,

the density operator, $TN(I)=N(I)$;

the density of kinetic energy operator: $TH_0(I)=H_0(I)$;

the density of momentum operator: $TP(I)=-P(I)$;

and the Liouville operator is time reversal invariant: $TLT=L$.

Instead of defining T on the *-Algebra R, the time reversal transformation could also be defined in the space of pure state $P_p$. For a pure state $\xi \in P_p$, the time reversal state $\xi_T = T\xi$ also belongs to $P_p$ with $(A, \xi_T) = (TA, \xi)$ for all $A \in R$. The pure states $\xi$ and $\xi_T$ correspond to pure state vectors $\rho$ and $\rho_T$, respectively. When $\rho$ has been transformed to $\rho_T$, all the infinitely many grid point state vectors of $\rho$ ought to be changed to their time reversal counterparts respectively.

Here we mainly focus on the discussions of spontaneous breaking of time reversal invariance in the dynamics of the SQINP. To meet this, only the quantum gas states will be considered, the states related to condensation phenomena will not be included in our state space P. All the pure state vectors in P fall into two categories. One contains only the time reversal invariant FPSV $\rho_0$ and all pure state vectors equivalent to $\rho_0$, i.e. all elements of the Hilbert Space $H_{\rho 0}$. Since the grid points are distinguishable, the time reversal invariant FPSV $\rho_0$ is the ground state vector of the system, in $\rho_0$ the grid point momentum density P(I) at any grid point should be zero. The GNS Construction $\{N_{\rho 0}, H_{\rho 0}\}$ associated with $\rho_0$ could be constructed. The $\rho_0$ and any other element $\rho_{0f}$ of $H_{\rho 0}$ have different grid point state vectors only at finitely many grid points. The time reversal counterpart $\rho_{0fT}$ of $\rho_{0f}$ is also inside the $H_{\rho 0}$. Hence, the dynamics of the QSINP is time reversal invariant in the GNS construction $\{N_{\rho 0}, H_{\rho 0}\}$. However, any element $\rho_{0f}$ of $H_{\rho 0}$ has grid point state vectors with non-zero momentum density P(I) only at finitely many grid points, but it has infinitely many grid point state vector with P(I)=0. Thus, all the state vectors of $H_{\rho 0}$ belong to the macroscopic state of 0K, and are macroscopically unrealizable. The remainders of the pure state vectors in P are called quantum gas pure state vectors, which involve grid point state vectors with non-zero momentum densities at infinitely many grid points. The quantum gas pure state vector $\rho_f$, and its time reversal counterpart $\rho_{fT}$ have different grid point state vectors at infinitely many grid points, thus they belong to two nonequivalent sets of pure state vectors. If $\rho_f$ belongs to the GNS Construction $\{N_\rho, H_\rho\}$ associated with the FPSV $\rho$, then $\rho_{fT}$ must belong to the GNS construction $\{N_{\rho T}, H_{\rho T}\}$ with the FPSV $\rho_T$. The $\{N_\rho, H_\rho\}$ and $\{N_{\rho T}, H_{\rho T}\}$ are non-equivalent. The time reversal transformation T is not the *inner-transformation* of GNS Construction associated with the quantum gas FPSV. T transforms the $\{N_\rho, H_\rho\}$ to another non-equivalent $\{N_{\rho T}, H_{\rho T}\}$, here $\rho_T = T\rho$. Inside $\{N_\rho, H_\rho\}$, the time reversal invariance breaks spontaneously, but the GNS Constructions $\{N_\rho, H_\rho\}$ and $\{N_{\rho T}, H_{\rho T}\}$ together compose a 2D representation of the time reversal group $\{T, I\}$.

As we have mentioned before, the whole dynamic process of the QSINP is within the GNS construction $\{N_\rho, H_\rho\}$ associated with initial FPSV $\rho$. Consequently, there is no dynamic means which can make the system to leap from one GNS Construction $\{N_\rho, H_\rho\}$ into another non-equivalent GNS construction $\{N_{\rho'}, H_{\rho'}\}$. From another point of view, the state vector $\rho_{A(t)}$ (with $\rho_A(t=0)=A\rho$) moves in $H_\rho$ according to equation (30), while the time reversal process $T\rho_A(t)$ (with $T\rho_A(t=0)=TA\rho$) moves in $H_{\rho T}$ which is disjointed with $H_\rho$. There are no dynamic methods that can make $\rho_A(t)$ into $H_{\rho T}$. Therefore, within the chosen GNS construction $\{N_\rho, H_\rho\}$, there exist only one dynamic direction of time: namely t>0. In such a sense, the time reversal invariance of dynamic laws is spontaneous breaking in the GNS construction $\{N_\rho, H_\rho\}$ associated with initial FPSV $\rho$ and thus the dynamic processes are time irreversible.

Brief summary, the *-Algebra R of the QSINP has infinitely many non-equivalent irreducible representations (GNS constructions). The dynamical motion of the QSINP is totally constrained in the GNS construction $\{N_\rho, H_\rho\}$ associated with the initial FPSV $\rho$, and time reversal transformation transforms one GNS construction $\{N_\rho, H_\rho\}$ to another non-equivalent irreducible GNS construction $\{N_{\rho T}, H_{\rho T}\}$. So, in each GNS construction the motion of the QSINP is time irreversible.

**IV Local Description and the Master Equation**

Different from the traditional quantum many body theory, not only the GNS Construction $\{N_\rho, H_\rho\}$ depends on the initial pure state vector $\rho$ and the motion of states are restricted in the sub-set of the space of pure state vector $H_\rho$, but also the Liouville equation (30) in $\{N_\rho, H_\rho\}$ only describes the time-irreversible motion. In $\{N_\rho, H_\rho\}$, the formal solution of the Liouville equation T(t)=Exp(-iLt) is a semi-group, but the time-irreversibility does not be explicitly shown by the expression (32), it is associated with GNS construction $\{N_\rho, H_\rho\}$ itself. In another way, even though there are the Hilbert space $H_\rho$, the C*-Algebras of linear operators $N_\rho$ on Hilbert space, and the Liouville equation (30) with the usual expression, we still could not quantitatively discuss the motion of the QSINP within our algebraic framework by some well-developed theoretical methods. In fact, it is necessary to incorporate the time irreversibility into the expression of time evolution operator T(t), i.e. to give the explicit expression of the semi-group.

For understanding the processes of approaching equilibrium of quantum many-body systems, Van Hove [13-14] and Prigogine *et. al*. [15-18] had discussed the long time asymptotic behavior of formal solution of Liouville equation for the fixed time direction t > 0, by using the quantum many-body theory (Fock space and quantum Liouville equation). In their studies, they assumed that the Liouville operator has continuous spectrum and therefore they could obtain a master equation with an explicit dissipative term. However the Liouville equation in Fock space is time reversible, it is unreasonable to discuss only the long time asymptotic behavior of one time direction. Furthermore, since the energy spectrum of the quantum system with finite number of particles is discrete, the formal scattering theory can't be applied, and hence, it is doubtful to get the generator of the semi-group by this way.

For the QSINP, it has already been noted that, the dynamical motion is time irreversible within the GNS Construction $\{N_\rho, H_\rho\}$ associated with the initial FPSV $\rho$. It is permitted to consider the long time asymptotic behavior of the solution of Liouville equation in the time direction t > 0. In addition, according to the spectrum theory of Hilbert space, the Liouville operator on $N_\rho$ is an boundless Hermitian operator, and thus, has real continuous spectrum along the whole real axis of complex plane, except the branch point z=0. It is reasonable to treat the Liouville equation in $\{N_\rho, H_\rho\}$ by the method of formal scattering theory.

In the GNS construction $\{N_\rho, H_\rho\}$, the Hamiltonian of the QSINP can be formally written as,

$$H = H_0 + H' = \sum_i^\infty \left( H_0 + \sum_j V \right)_i , \qquad (33)$$

where *i* sum over all infinitely many grid points and j over finitely many grid points close to grid point *i*. We should note that none of H, $H_0$ and H′ are the elements of C*- algebra $N_\rho$, but each term in their formal summation of *i* should belong to $N_\rho$. Because the Liouville operator L is a Hermitian operator on $N_\rho$, the time evolution operator T(t) of the Liouville equation in $\{N_\rho, H_\rho\}$, could still be defined as a divergent infinite series by equation (33).

The resolvent operator could be introduced:

$$R(z) = 1/(z-L), \qquad (34)$$

where z is the complex variable. Since the spectrum of L lie on the real axis of complex z plane, and only the time direction t>0 should be considered in $\{N_\rho, H_\rho\}$, we have.

$$T(t) = \int_{C+i0} \text{Exp}(-izt) R(z) dz . \qquad (35)$$

Here the integral circuit C+i0 is shown in Fig.1.

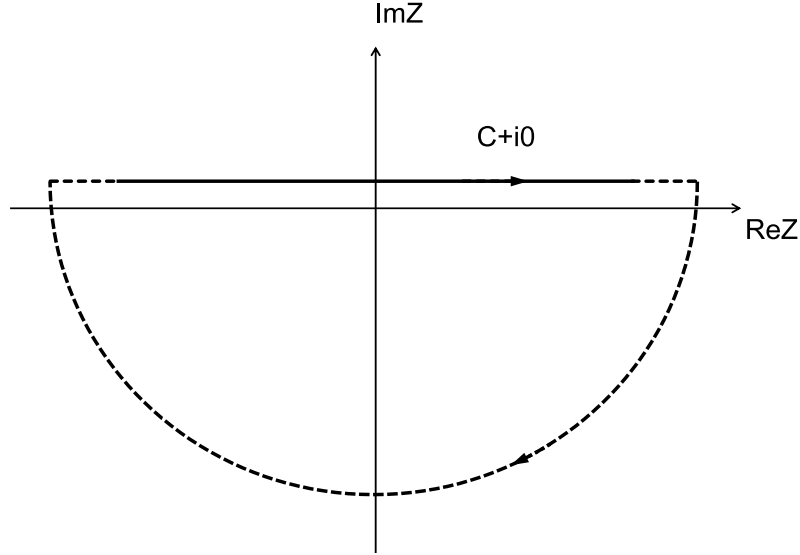

Fig.1 The integral circuit of equation (34) in complex plan z.

Though the Liouville equation formally describes the motion of the infinite degrees of freedom, only the motion of a few degrees of freedom under the influences of others could be computable. Such a few degrees of freedom are usually related to the macroscopic observables of the QSINP and compose a subspace $R_S$ in R. By introducing a projection operator P on $N_\rho$, which can project any element of R into the subspace $R_S$, and Q=1-P. P and Q satisfy the relations: $P^2=P$; $Q^2=Q$; and A=PA+QA, $A \in N_\rho (\text{or } R)$. The operators on $N_\rho$ (or R) are divided into four parts as followings,

L=PLP+PLQ+QLP+QLQ ;

T(t)=PT(t)P+PT(t)Q+QT(t)P+QT(t)Q ;

R(z)=PR(z)P+PR(z)Q+QR(z)P+QR(z)Q

From equation (35), we have,

$$PR(z)P = (1/(z-PLP-E(z)))P \qquad (36)$$

and,

$$E(z) = PLQ(1/(z-QL \qquad (37)$$

E(z) can be regarded as the corrections to the denominator of PR(z)P, given rise by the *interaction* terms PLQ, QLP and QLQ, respectively.

Since there is only one Hermitian operator L both in R(z) and T(t), it is possible to treat L in the integrand of integration (36) as the set of all eigenvalues of L on the real axis. For a system with finitely many particles, L has a discrete spectrum distributed on the real axis of complex plane z. No matter how dense the spectrum points are, the integration circuit can't be analytically continued to the lower half of the second Riemann surface. So, it is necessary to calculate all the contributions of these discrete spectrum points, of course, such calculations are as complicate as solving the full Liouville equation. But for the QSINP, L is the operator on $N_\rho$, its resolvent should be defined by the divergent infinite series (35) and is boundless for non-zero real z values. According to the theory of Hilbert space, the spectrum of L is continuous on the whole real axis except the origin z = 0. The real axis in complex plane z is the secant line of the integrand in integration (36) with branch point at z=0. In complex function theory, the result of integration (36) is contributed by the residues of the poles in the lower half of the second Riemann surface. These poles are determined by the equation,

$$z - PLP - E(\ ) \tag{38}$$

For determining the motion of the QSINP at macroscopic time scale, to discuss the long time asymptotic behavior of T(t) with t>0 is necessary. Hence only the residue of the pole $z_0=PLP+E(z_0)$, which is closest to the point z=0, should be considered. In weakly coupling cases, this pole is determined approximately by,

$$z_0 = PLP + E(-i0) = PLP + PLQ(1\ (- \tag{39}$$

The time evolution operator becomes,

$$T_P(t) = PT(t)P = \mathrm{Exp}(-iz_0 t) \tag{40}$$

By now, $T_P(t)$ and $z_0$ in equation (39) and (40) are the operators defined on $N_\rho$, where $N_\rho$ is the set of all linear operators on the Hilbert space $H_\rho$ associated with the initial FPSV $\rho$. But the different $N_\rho$ of non-equivalent GNS Constructions are isomorphic to R. So in Heisenberg Picture, the Liouville operator L、the projection operators P and Q could be regarded as linear operators on R. The time evolution operators (39) and (40) do not depend on the GNS construction $\{N_\rho, H_\rho\}$ (or the initial fully pure state vector $\rho$) explicitly, thus, they could be taken as the evolution operators on the *- algebra R.

If the processes are dissipative, the *dissipation condition* is,

$$z_0 = PLP + E(-i0)$$
$$= PLP + PLQ(1/(-i0-QLQ))QLP, \quad (41)$$

has a non-positive imaginary part. Let $z_0 = \xi - i\theta$, $\xi$ and $\theta$ are both Hermitian operators,

$$\xi = PLP + PLQp(1/QLQ)QLP \quad (42)$$

$$\theta = PLQ\delta(QLQ)QLP \quad (43)$$

where P means the main value of the integration, $\delta$ function $\delta(QLQ)$ indicates energy conservation for the process QLQ. The time evolution operator of subspace P can be obtained,

$$T_P(t) = \text{Exp}(-i\xi t - \theta t) \quad (44)$$

The operator $\{T_P(t), t>0\}$ is a semi-group. The Hermitian operator $\xi$ describes the dynamic motion of macroscopic degrees of freedom and can be called as the *dispersion operator*. It is easy to prove that the Hermitian operator $\theta$ is non-negative and has a zero eigenvalue, the eigenfunction of which corresponds to the equilibrium state. The operator $\theta$ describes the process of approaching equilibrium. Equation (44) can be rewritten into the form of master equation:

$$dT_P(t)/dt = (-i\xi - \theta)T_P(t). \quad (45)$$

The master equation (45) is an initial state independent operator equation on the *-algebra R. The selection of the project operator P, strictly speaking, depends on what kind of macroscopic process we would like to study, such as, for studying the heat conduction of gas, the projection operator P should be related with the energy current operators. Since the number of macroscopic parameters is too smaller compared with the number of microscopic degrees of freedom, the different projection operators, which are related with different macroscopic processes respectively, couldn't lead to significantly different master equations. Such possibilities illustrate the typical links between microscopic and macroscopic descriptions of system with infinitely many degrees of freedom.

**V. Conclusions and Remarks**

In summary, there are two kinds of methods to discuss the motion of the quantum system with enough many particles. First one is that, the motions of finitely many particles were studied, and then the thermodynamic limit was taken. Throughout the whole treatment, the observers are always outside (or above) the system. Such method belongs to the framework of Foke space theory. In our paper, starting directly from the quantum system with infinite number of particle, the motions of the finitely many particles on the background of infinitely many particles has been discussed. The finitely many particles could be few or as many as possible, but the surrounding background of infinitely many particles always exists. This is the basic difference of our way with the ways of thermodynamic limit. It could be thought that, just due to the existence of such infinite *background*, the finite dynamic operations could not realize the time reverse of the whole system. In our theory the observers are inside the system, they could not observe the boundary, or outside, of the system and so they can only discuss the local macroscopic properties of the system, such as the local thermo-conductivity, diffusion coefficient.

Strictly speaking, only the finite system could be described quantitatively. The quantum system with infinite number of particle isn't measurable as a whole, and the defined dynamic variable algebra R is a *-Algebra. But due to the fact that, any element of R is only related with finitely many degrees of freedom, the well-known GNS Construction in C*-Algebra theory naturally could be the link between the infinite system with description of local finite sub-system. Here the word *local* does not mean the isolated local subsystem with finite size, but the subspace in state vector space. By this way of thinking, the spontaneously breaking of the time reversal invariance in the GNS construction associated with initial fully pure state vector could easily be understood.

In all, the way directly to treat the dynamic problems of the quantum system with infinite number of particles has been developed and it serves as a bridge to link the macroscopic and the microscopic dynamic depictions of quantum gas system.

**Acknowledgement**: I acknowledge to Prof. Berlin Hao, Prof. Lu Yu, Prof. Gongqing Zhang, Prof. Melin Ge, and Prof. Ke Wu for their valuable discussions. I am also grateful to Prof. Yu Jia, Prof. Fei Wang, Prof. Yigang Cao and Prof. Qi Zhang at Zhengzhou University for their valuable supports. This work was partly supported by the NSF of China(Grant. No. 11071126), and partly by the "973" project of China.